\documentclass[pdflatex,sn-nature]{sn-jnl}

\usepackage{graphicx}%
\usepackage{multirow}%
\usepackage{amsmath,amssymb,amsfonts}%
\usepackage{amsthm}%
\usepackage{mathrsfs}%
\usepackage[title]{appendix}%
\usepackage{xcolor}%
\usepackage{textcomp}%
\usepackage{manyfoot}%
\usepackage{booktabs}%
\usepackage{algorithm}%
\usepackage{algorithmicx}%
\usepackage{algpseudocode}%
\usepackage{listings}%
\usepackage{xcolor}

\theoremstyle{thmstyleone}%
\theoremstyle{thmstyletwo}%

\theoremstyle{thmstylethree}%

\begin{document}

\title[Article Title]{Interfacial control of hot-carrier extraction and photostability in two-dimensional materials}



\author[1,2]{\fnm{Claudia} \sur{Gollner}}\email{cgollner@stanford.edu}

\author[3]{\fnm{Mohammad} \sur{Taghinejad}}\email{mtaghi@stanford.edu}

\author[3]{\fnm{Chenyi} \sur{Xia}}\email{chenyix@stanford.edu}

\author[3]{\fnm{Zhepeng} \sur{Zhang}}\email{zhangzp@stanford.edu}

\author[4]{\fnm{Fang} \sur{Liu}}\email{fliu10@stanford.edu}

\author[5]{\fnm{Francesco} \sur{Laudani}}\email{francesco.laudani@tuwien.ac.at}

\author[5]{\fnm{Annette} \sur{Foelske}}\email{annette.foelske@tuwien.ac.at}
 
\author[3,6]{\fnm{Mark L.} \sur{Brongersma}}\email{markb29@stanford.edu}

\author[3]{\fnm{Andrew J.} \sur{Mannix}}\email{ajmannix@stanford.edu}

\author[1,2]{\fnm{Tony F.} \sur{Heinz}}\email{tony.heinz@stanford.edu}

\author*[2,3]{\fnm{Aaron} \sur{Lindenberg}}\email{aaronl@stanford.edu}

\affil[1]{\orgdiv{Department of Applied Physics}, \orgname{Stanford University}, \orgaddress{\street{348 Via Pueblo}, \city{Stanford}, \postcode{94305}, \state{CA}, \country{USA}}}

\affil[2]{\orgdiv{Stanford Institute for Materials and Energy Sciences}, \orgname{SLAC National Accelerator Laboratory}, \orgaddress{\street{2575 Sandhill Rd}, \city{Menlo Park}, \postcode{94025}, \state{CA}, \country{USA}}}

\affil[3]{\orgdiv{Department of Materials Science and Engineering}, \orgname{Stanford University}, \orgaddress{\street{496 Lomita Mall}, \city{Stanford}, \postcode{94305}, \state{CA}, \country{USA}}}

\affil[4]{\orgdiv{Department of Chemistry}, \orgname{Stanford University}, \orgaddress{\street{333 Campus Drive}, \city{Stanford}, \postcode{94305}, \state{CA}, \country{USA}}}

\affil[5]{\orgdiv{AIC - Analytical Instrumentation Center}, \orgname{TU Vienna}, \orgaddress{\street{Lehargasse 6}, \city{Vienna}, \postcode{1060}, \country{Austria}}}

\affil[6]{\orgdiv{Geballe Laboratory for Advanced Materials}, \orgname{Stanford University}, \orgaddress{\street{496 Lomita Mall}, \city{Stanford}, \postcode{94305}, \state{CA}, \country{USA}}}


\abstract{
Two-dimensional transition metal dichalcogenides (TMDCs) are promising materials for next-generation optoelectronic devices, yet their implementation is hindered by limited sample stability and challenges in forming reliable electrical contacts. 
Here, by utilizing time-domain THz emission spectroscopy we directly probe charge carrier dynamics in monolayer  WS\textsubscript{2} on gold (Au) and fused silica (SiO\textsubscript{2}) as a function of interface morphology.
For laser excitation above the band gap of WS\textsubscript{2}, we independently extract effective transport times for both electrons and holes and find that discontinuous WS\textsubscript{2} contacts on rough Au generate larger net photocurrents than uniform, strongly coupled interfaces \textemdash \ a counterintuitive observation attributed to imbalanced electron and hole transfer from WS\textsubscript{2} to Au. 
Crucially, we demonstrate that ultrafast charge extraction and separation suppress recombination-driven energy release and thereby prevent photo-induced degradation under ambient conditions, eliminating the need for encapsulation. These findings redefine interfacial design as a central control parameter for both performance and stability in 2D optoelectronic devices.
}

\keywords{2D transition metal dichalcogenides, ultrafast charge carrier dynamics, interface morphology, photo-induced degradation}



\maketitle

\section{Main}\label{sec1}
Two-dimensional transition metal dichalcogenides are model systems for exploring novel physics and enabling next-generation optoelectronics, owing to their atomically thin structure, tunable band gaps, high carrier mobility, and strong interlayer interactions in van der Waals (vdW) heterostructures \cite{Wang:2012, Schmidt:2015, Liao:2020, Schaibley:2016}.
Despite this promise, practical device integration remains challenging. 
Key limitations include degradation of the atomically thin layers in ambient environment \cite{Kang:2017,Gao:2016, Ly:2014},  high contact resistance arising from large Schottky barrier heights (SBHs) and uncontrollable Fermi-level pinning (FLP) \cite{Liu:2025, Xiaochi:2022, Wang_Yan:2019}.
These bottlenecks encourage a better understanding of interfacial charge injection and the underlying nature of deterioration at metal-TMDC contacts, as these ultrafast, atomic-scale processes ultimately define device efficiency and reliability.
Interfacial carrier dynamics also play a pivotal role in technologies that seek to extract hot photocarriers before they dissipate energy, thereby enabling their direct use in optoelectronic devices.
Despite their importance, studies on charge carrier dynamics at TMDC-metal interfaces remain scarce.   
Most investigations rely on transient absorption spectroscopy \cite{Shan:2019,Hong:2025}, probing charge transfer indirectly through carrier lifetimes, or time- and angle-resolved photoemission spectroscopy (TR-ARPES)  \cite{Cabo:2015,Ulstrup:2017}, which offers direct insights but requires ultra-high vacuum that  does not reflect realistic device environments.\\
Here, we employ time-domain terahertz (THz) emission spectroscopy (TES) as a powerful method to directly probe interfacial charge transfer dynamics at the interface of monolayer (ML) $\mathrm{WS_2}$ on Au or  $\mathrm{SiO_2}$. 
We find that TMDC/metal junctions with smooth interfaces and strong electronic coupling, accompanied by orbital hybridization, exhibit significantly faster and more efficient interfacial charge transfer. 
Counterintuitively, for above-band gap excitation, we show that discontinuous contacts on rough metal substrates generate enhanced, hole-dominated photocurrents due to asymmetric charge transfer across the interface. 
In turn, uniform interfaces yield a fast but balanced carrier flow, thereby significantly reducing the net current, as confirmed by our theoretical model.\\
Enabling a carrier transport channel between TMDC MLs and a surrounding material is further shown to suppress photo-chemical degradation in ambient conditions.
On insulating $\mathrm{SiO_2}$, excited charge carriers are confined within the ML.
Their recombination triggers defect-mediated reactions with atmospheric $\mathrm{H_2O}$ and $\mathrm{O_2}$.
Conversely, the $\mathrm{WS_2}$/Au interface facilitates ultrafast (sub-100 fs) carrier extraction that outpaces these picosecond to nanosecond-scale degradation pathways, ensuring material stability.
These insights establish interface engineering as a critical framework for optimizing both ultrafast charge-carrier extraction and long-term environmental stability in ML optoelectronics.
\section{Metal -TMDC interface morphology}\label{sec2}
\begin{figure}[h]
\centering
\includegraphics[width=1\textwidth]{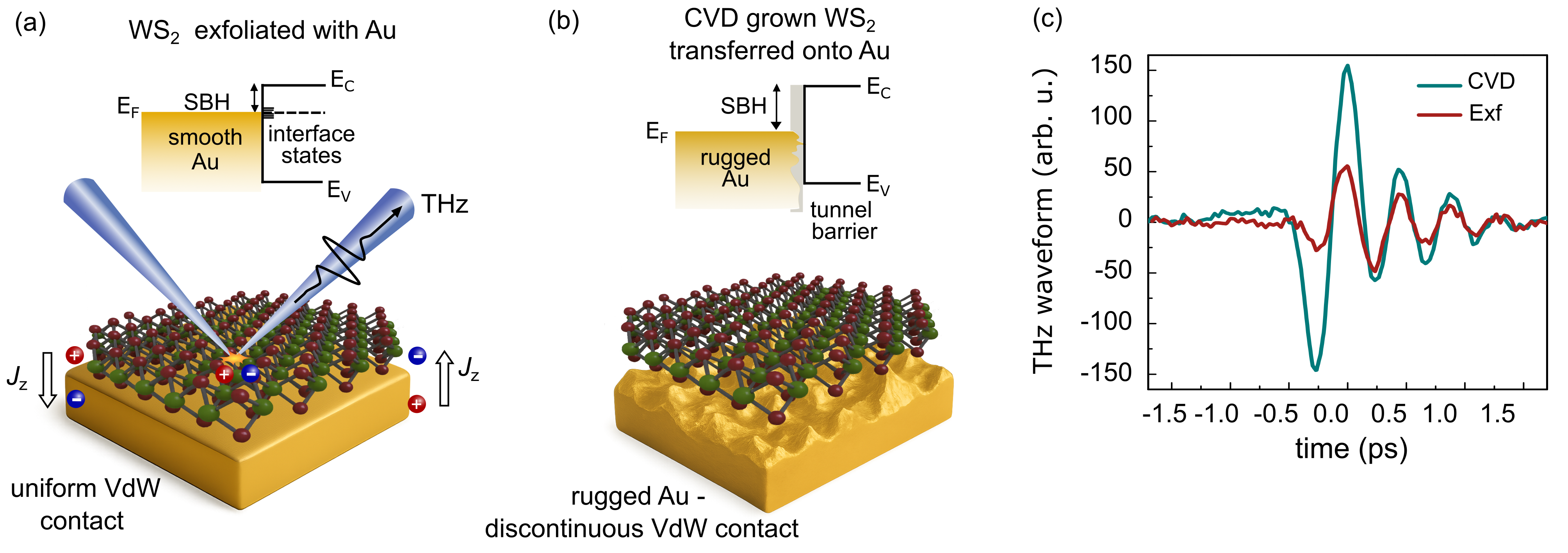}
\caption{ 
(a)-(b) top: Exaggerated energy band diagram at the interface, including interface gap states for Exf - $\mathrm{WS_2}$, highlighting a reduced Schottky barrier height (SBH) for Exf - $\mathrm{WS_2}$ and tunnel barrier for CVD-$\mathrm{WS_2}$
(a)-(b) bottom: Schematic illustrations of exfoliated  and CVD-grown  $\mathrm{WS_2}$ on Au, showing a flush, uniform vdW contact for Exf-$\mathrm{WS_2}$ and a discontinuous contact on the rough Au surface for CVD-$\mathrm{WS_2}$. 
Panel (a) also includes a schematic of interfacial currents $J_{\mathrm{z}}$ in the $\mathrm{WS_2}$/Au stack that enables transient THz pulse emission. 
(c) THz waveforms from Exf- $\mathrm{WS_2}$ (red) and CVD-$\mathrm{WS_2}$ (green) following $3.12\ \mathrm{eV}$ femtosecond excitation. Despite an additional tunnel barrier arising from the poor metal-TMDC contact,  CVD-$\mathrm{WS_2}$ generates a stronger THz signal.
}\label{fig1}
\end{figure}

To probe the influence of interface morphology on the energy band alignment and charge carrier dynamics, we prepared $\mathrm{WS_2}$ MLs on Au using two approaches: (i) direct exfoliation from bulk crystals with a template-stripped Au substrate \cite{Fang:2020}, yielding atomically flat, contamination-free interfaces, and (ii) hybrid metal-organic chemical vapour deposition (CVD) \cite{Zhepeng:2024} on Sapphire followed by PMMA-assisted wet transfer onto a thermally evaporated Au substrate (hereafter labelled as Exf-$\mathrm{WS_2}$ and CVD-$\mathrm{WS_2}$, respectively). 
Details of the preparation procedure can be found in the method section.
Exf-$\mathrm{WS_2}$ exhibits an ultra-smooth surface with an average roughness of $\mathrm{R_a} = 0.251\ \mathrm{nm}$ (see AFM images in Supplementary Fig. S1) and a uniform vdW contact.
In contrast, CVD-$\mathrm{WS_2}$ rests partially suspended on a rugged Au surface, exhibiting a surface roughness of $\mathrm{R_a} =  1.06\ \mathrm{nm}$.
We further employed Kelvin probe force microscopy (KPFM), photoluminescence (PL) and Raman spectroscopy to characterise the interface, revealing doping type and differences in band alignment. 
Schematic illustrations of the observed interface morphology and energy band structure of Exf-$\mathrm{WS_2}$ and CVD-$\mathrm{WS_2}$ are shown in Fig. \ref{fig1}(a) and (b), respectively.
Exf-$\mathrm{WS_2}$ exhibits a higher surface potential (see KPFM images in Supplementary Fig. S1) than CVD-$\mathrm{WS_2}$, indicating stronger electron transfer from $\mathrm{WS_2}$ to Au and the formation of interface states reducing the SBH. 
For the CVD-$\mathrm{WS_2}$ stack, the work function differs from bare Au by only $4\ \mathrm{mV}$ (as compared to $21\ \mathrm{mV}$ for Exf-$\mathrm{WS_2}$), consistent with weak interaction and low charge transfer, approaching the Schottky-Mott limit.  
Assuming the same energy band position with respect to vacuum-level, both $\mathrm{WS_2}$-Au stacks are relatively n-doped, with a higher doping level for Exf-$\mathrm{WS_2}$.
Raman measurements of Exf-$\mathrm{WS_2}$ reveal a splitting of the out-of-plane phonon mode (Supplementary Fig. S1), indicating a weakening of the W-S bond caused by orbital hybridization between sulfur atoms and the contact metal  \cite{Pollmann:2021, Gong:2014}. 
Although both samples possess an intrinsic tunneling gap arising from vdW stacking, our results indicate a larger effective tunneling barrier in CVD samples due to incomplete metal-TMDC contact and possible molecular contamination at the interface. \\
Interestingly, despite the strong coupling of Exf-$\mathrm{WS_2}$ on Au, above band gap excitation with 3.12 eV fs-pulses reveal a larger net interfacial photocurrent in CVD-$\mathrm{WS_2}$, directly observed by the emitted THz field shown in Fig. \ref{fig1}(c), detected by free-space electro-optic sampling (EOS) with a $1\ \mathrm{mm}$ ZnTe crystal. 
By capturing the temporal and polarization characteristics of the radiated electric fields, TES offers a contact-free, quantitative method to determine the magnitude, direction, and temporal evolution of the photocurrent \cite{Gollner:2026}. 
Previous works on charge carrier dynamics in TMDC-metal interfaces based on transient absorption measurements have been unable to distinguish between electron and hole transfer processes, as interlayer dynamics are indirectly extracted from the lifetime of excited states  \cite{Hong:2025}.
Consequently, the counterintuitive observation of an enhanced transient photocurrent density for a reduced interface contact quality has not been discussed so far.
Because gold strongly absorbs light with photon energies near $3\ \mathrm{eV}$, the photo-induced interfacial current \ \textemdash \ driven by carriers in both the Au film and TMDC layer \ \textemdash \ can evolve in both out-of-plane directions, as indicated by the photocurrent density $J_{\mathrm{z}}$ arrow in Fig.\ref{fig1}(a).
To elucidate details on charge carrier dynamics, we first examine TES for the simple case of below-band gap excitation, excluding charge transfer due to photo-excited carriers from $\mathrm{WS_2}$ to Au.
We further carefully sample the emitted THz field for below- and above band gap excitation with a thin $250 \ \mathrm{\mu m}$ GaP crystal allowing broad bandwidth detection, as described in the following sections, shedding light on charge transfer processes for different interface morphologies.   
\section{Hot electron injection to $\mathrm{WS_2}$ for below- band gap excitation}\label{sec3}
\begin{figure}[h]
\centering
\includegraphics[width=1\textwidth]{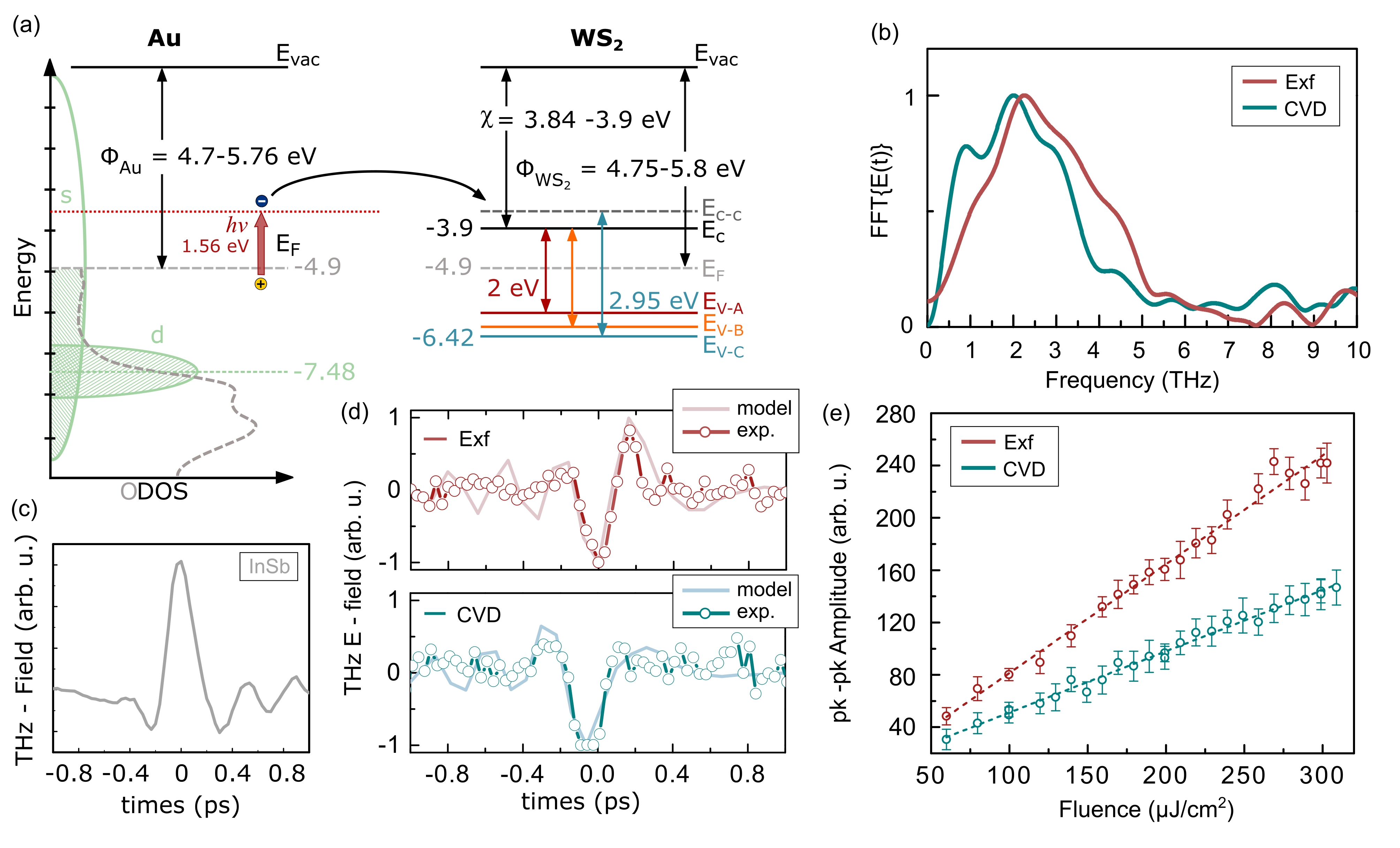}
\caption{(a) Energy band alignment of Au (left) and monolayer $\mathrm{WS_2}$ (right) before contact. Reported Au work functions are taken from \cite{Gong:2014, Markeev:2021, Krolikowski:1970, Tang:2019}, with electronic density of states (DOS, green) from \cite{Dreesen:2002} and optical density of states (ODOS, grey dashed) from \cite{Krolikowski:1970}. $\mathrm{WS_2}$ conduction and valence band parameters are taken from \cite{Eroglu:2021,Ansari:2018, Li:2014}, electron affinity $\chi$ \cite{Mohl:2020, Kang:2013, Frisenda:2018, Liu:2016} and work function $\Phi_{\mathrm{WS_2}}$ \cite{Markeev:2021, Hill:2016}. 
Excitation with $1.56\ \mathrm{eV}$ pulses predominantly excites free electrons just below the Au Fermi level $\mathrm{E_F}$, providing sufficient excess energy for interfacial transfer into $\mathrm{WS_2}$.
(b) Fourier spectra of the time-domain signals shown in (d) for Exf-$\mathrm{WS_2}$ (red) and CVD-$\mathrm{WS_2}$ (green).
(c) THz field from reference InSb.
(d) THz waveforms from Exf-$\mathrm{WS_2}$ (red dotted line) and CVD-$\mathrm{WS_2}$ (green dotted line), measured with a $250\ \mathrm{\mu m}$-thick GaP crystal, together with the corresponding waveform from our charge transport model, demonstrating faster charge transfer for Exf-$\mathrm{WS_2}$.
(e) Peak-to-peak THz amplitude versus pump fluence for Exf-$\mathrm{WS_2}$ (red) and CVD-$\mathrm{WS_2}$ (green) as a measure of the transient photocurrent.}\label{fig2}
\end{figure}

The spatiotemporal dynamics of interfacial charge transfer are governed by photocarrier energy and band alignment, necessitating a quantitative description of the interfacial energy landscape to understand transfer probabilities.
Figure \ref{fig2}(a) illustrates the isolated energy diagram of Au  (left) and ML $\mathrm{WS_2}$ (right) before the formation of a junction. 
Reported work functions of Au span from $\Phi_{\mathrm{Au}}=4.7-5.76\  \mathrm{eV}$ \cite{Gong:2014, Markeev:2021, Krolikowski:1970, Tang:2019}, and $\Phi_{\mathrm{WS_2}}=4.75-5.8\  \mathrm{eV}$ \cite{Markeev:2021, Hill:2016} for $\mathrm{WS_2}$, depending on substrate and growth method.
Au exhibits a high density of states (DOS) from filled 5\textit{d} orbitals around $2.58\ \mathrm{eV}$ below the Fermi level and free electron states from the partially filled \textit{sp}-like conduction band, as presented by the green dashed area in Fig. \ref{fig2}(a), adapted from ref \cite{Dreesen:2002}.
The dashed grey line represents the corresponding optical density of states (ODOS), as calculated in ref \cite{Krolikowski:1970}, with a pronounced absorption maximum at $3\ \mathrm{eV}$. 
In $\mathrm{WS_2}$, A- and B- exciton resonances around $2\ \mathrm{eV}$ \cite{Ansari:2018, Li:2014} correspond to direct transitions at the \textit{K}-valley  with an electron affinity of $\chi = 3.84-3.9\  \mathrm{eV}$ \cite{Mohl:2020, Kang:2013, Frisenda:2018, Liu:2016}.
The nested C-exciton transition of 2.95 eV  \cite{Eroglu:2021} \ \textemdash \ featuring band offsets of 0.52 eV and 0.33 eV relative to the valence band maximum and conduction band minimum of the A exciton, respectively \cite{Feng:2022} \ \textemdash \  has a valance band maximum in close proximity to the energy range of the Au \textit{d}-orbitals.
Illumination with 1.56 eV pulses, having photon energies below the band gap of $\mathrm{WS_2}$, primarily excites free conduction electrons in Au. 
Only hot electrons originating from just below the Fermi level possess sufficient energy to overcome the Au/$\mathrm{WS_2}$  interfacial energy barrier and generate a transient current pulse that emits THz radiation.
Note that, according to KPFM measurements, this energy barrier or SBH is smaller for exfoliated samples, thereby increasing the probability of hot electron transfer to $\mathrm{WS_2}$.
Detection of fast carrier dynamics is facilitated with a  $250 \ \mathrm{\mu m}$ thick GaP crystal.
The THz signal depicted in Fig. \ref{fig2}(d)  emerging from the Exf-$\mathrm{WS_2}$ sample (red) exhibits a shorter time evolution and corresponding broader Fourier spectrum ($\mathrm{FWHM} \simeq 3.38 \ \mathrm{THz}$, see Fig.\ref{fig2}(b)) compared to CVD-$\mathrm{WS_2}$ ($\mathrm{FWHM} \simeq 2.97 \ \mathrm{THz}$, green), demonstrating faster interfacial charge transfer. 
Fig. \ref{fig2}(c) shows a reference THz waveform from InSb measured under identical conditions, which is known to emit THz radiation due to the photo-Dember effect with electrons diffusing faster into the bulk material. 
The opposite polarity of the THz waveforms observed from $\mathrm{WS_2}$/Au confirms hot-electron injection from Au into $\mathrm{WS_2}$ as the dominant transient photocurrent. 
We model the THz emission using coupled rate equations describing the density of (hot) electrons in Au and $\mathrm{WS_2}$ (see Section 6 of the Supplementary Information).
Only ballistic, nonthermal electrons in Au contribute to THz generation in the model, consistent with the experimentally observed linear scaling of the THz field amplitude with excitation fluence (see Fig. \ref{fig2}(e)). 
Electrons injected into the conduction band of $\mathrm{WS_2}$ either return directly to Au or first undergo intraband relaxation before returning. Following previous reports \cite{Chen:2016, Ma:2019}, the latter pathway is assumed to play a minor role.
The resulting carrier dynamics are captured by an effective retraction rate $\gamma_{\mathrm{r}}$, which quantifies charge transfer from $\mathrm{WS_2}$ back to Au and is governed by the interfacial band alignment and interface morphology, as discussed above. 
The best agreement with experiments is obtained for $\gamma_{\mathrm{r}}^{\mathrm{Exf}} = 1/50\ \mathrm{fs^{-1}}$ for Exf-$\mathrm{WS_2}$ and $\gamma_{\mathrm{r}}^{\mathrm{CVD}} = 1/160\ \mathrm{fs^{-1}}$ for CVD-$\mathrm{WS_2}$.
Peak-to-peak THz amplitudes as a function of pump fluence  shown in Fig. \ref{fig2}(e) are measured using a 1 mm ZnTe crystal. 
After correcting the detected amplitudes for missing spectral components (Supplementary Fig. S3), Exf-$\mathrm{WS_2}$ exhibits a larger THz signal. 
Because the THz field scales with the time derivative of the transient photocurrent, the larger peak amplitude reflects the significantly shorter transient photocurrent.
From a quantum mechanical perspective, carrier injection is governed by the interlayer tunneling rate, which is enhanced by strong orbital overlap, reduced interface spacing and energy barrier height. 
The larger THz signal observed for Exf-$\mathrm{WS_2}$ confirms this picture, demonstrating improved charge transfer efficiency when photocurrents originate solely from hot electrons in the metal with strong interface coupling.
\section{Hot carrier injection under above-band gap excitation}\label{sec3}
\begin{figure}[h]
\centering
\includegraphics[width=1\textwidth]{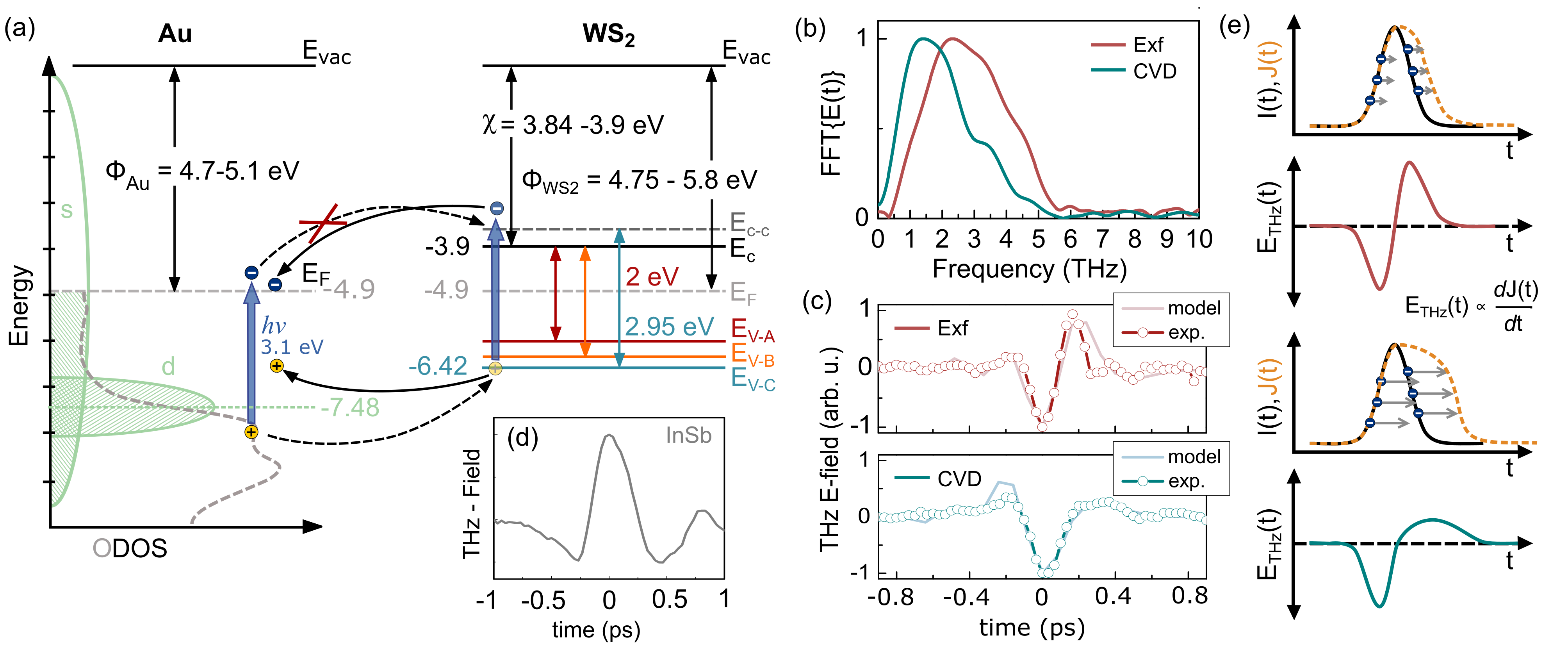}
\caption{(a) Schematic energy band diagram showing simultaneous electron and hole extraction from $\mathrm{WS_2}$ to Au following above-band gap excitation. 
(b) Fourier spectra of the THz waveforms in (c), highlighting the broader bandwidth for Exf-$\mathrm{WS_2}$ (red), indicative of faster charge transfer.
(c) THz fields measured with a $150\ \mathrm{\mu m}$ GaP crystal in dry air for Exf-$\mathrm{WS_2}$ (red dotted line) and CVD-$\mathrm{WS_2}$ (green dotted line), and their corresponding modeled THz waveforms.
(d) Reference THz pulse from InSb, confirming hole injection into Au as the dominant photocurrent contribution.  
(e) Conceptual link between THz waveform shape and carrier transit time: the excitation pulse intensity envelope $I(t)$  (black) determines the rise of the photocurrent  $J(t)$ (orange dashed), while its decay reflects the interfacial transit time. Because the THz field is proportional to the time derivative of the photocurrent density, fast transfer yields a symmetric, pulse-limited waveform (red), whereas slower transfer produces a stretched, asymmetric waveform (green).}\label{fig3}
\end{figure}

The fundamental principle of optoelectronic devices, such as photovoltaics or photodetectors, is to generate excited electron-hole pairs and  either harness their energy or extract charge carriers to utilize them in an external circuit. 
As such, it is of crucial importance to understand charge carrier dynamics when the photon energy exceeds the energy band gap of the semiconductor. 
For above-band gap excitation, the $795\ \mathrm{nm}$ laser pulse was frequency-doubled using a $0.5\ \mathrm{mm}$ thick BBO crystal, yielding $3.12\ \mathrm{eV}$ pulses with a duration of $\sim 76\ \mathrm{fs}$ (accounting for group velocity dispersion). 
Such high-energy photons primarily interact with \textit{d}-band electrons in Au that are located deep under $E_{\mathrm{F}}$, creating hot holes and free electrons with energies slightly above $E_{\mathrm{F}}$, unable to transfer into $\mathrm{WS_2}$.
In $\mathrm{WS_2}$, $3.12\ \mathrm{eV}$ photons are strongly absorbed  at the C-exciton transition (see Fig. \ref{fig3}(a)), generating hot electrons and holes in the ML with sufficient excess energy to transition into the Au substrate. 
Although hot holes excited in Au are energetically capable of transferring into $\mathrm{WS_2}$, the much higher density of states in 3D Au makes hot-hole injection from $\mathrm{WS_2}$ into Au the more likely process.
Figure \ref{fig3}(c) shows the THz waveforms from different $\mathrm{WS_2}$/Au interface morphologies, measured with a $150\ \mathrm{\mu m}$ GaP crystal, together with their modeled THz waveforms. 
The opposite polarity of the THz signal from InSb shown in \ref{fig3}(d) and the $\mathrm{WS_2}$/Au samples indicates oppositely directed net currents.
Because THz emission in InSb arises from photoelectron diffusion into the bulk, THz generation in $\mathrm{WS_2}$/Au must result either from a net transfer of electrons from Au to $\mathrm{WS_2}$  or from a net transfer of photoexcited holes from $\mathrm{WS_2}$  to Au. 
As most  photoelectrons excited in Au by 3.12 eV photons lack sufficient energy to overcome the interfacial barrier, the observed THz emission is attributed to a net hole flux from $\mathrm{WS_2}$ into the Au substrate.
The THz spectra (Fig. \ref{fig3}(b)) reveal a blue shifted and broader bandwidth for exfoliated $\mathrm{WS_2}$ ($\mathrm{FWHM} \simeq 3.24\ \mathrm{THz}$) compared to CVD-$\mathrm{WS_2}$ ($\mathrm{FWHM} \simeq 2.23\ \mathrm{THz}$), consistent with faster charge transfer. 
The nearly symmetric waveform of  Exf-$\mathrm{WS_2}$ suggests that the charge transfer time is limited by the duration of the excitation laser pulse (see sketch in Fig. \ref{fig3}(e)), indicating that the intrinsic charge transfer time is significantly shorter than the pump pulse duration.
\begin{figure}[h]
\centering
\includegraphics[width=1\textwidth]{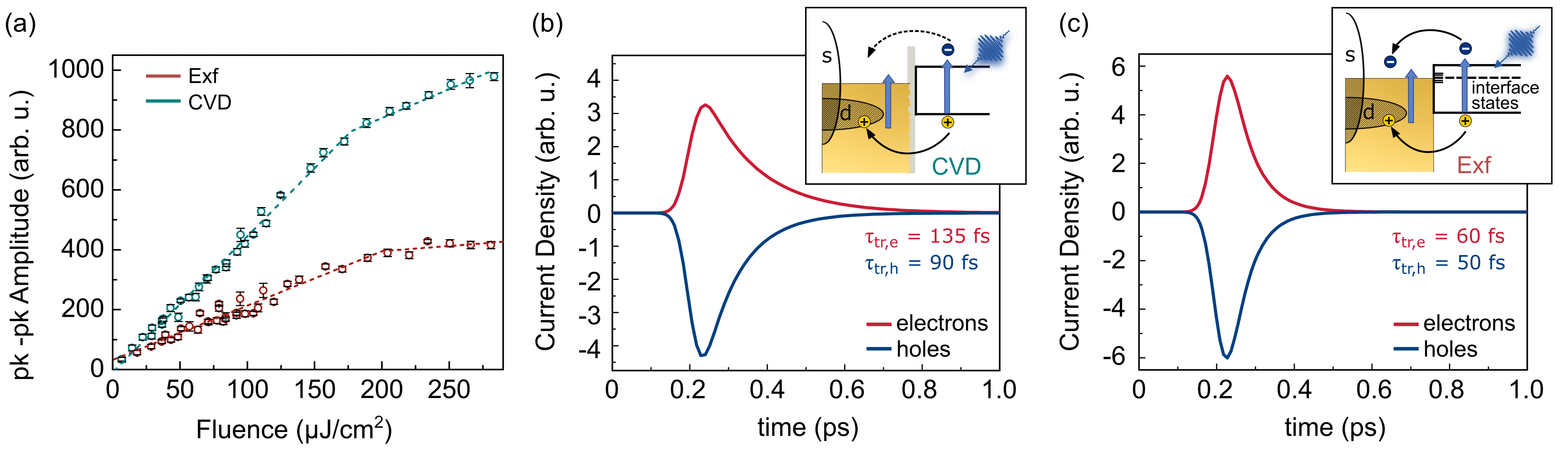}
\caption{(a) Peak to peak THz amplitude for Exf-$\mathrm{WS_2}$ (red) and CVD-$\mathrm{WS_2}$ (green) measured in dry air with a $1\ \mathrm{mm}$ ZnTe crystal as a function of pump fluence.
(b)-(c) Modelled transient current densities for electrons (red) and holes (blue), along with the corresponding transport times from $\mathrm{WS_2}$ to Au, demonstrating asymmetric electron and hole transfer in  CVD-$\mathrm{WS_2}$ (b) and efficient but  balanced transfer in exfoliated $\mathrm{WS_2}$ (c), resulting in a diminished net photocurrent.
Insets illustrate schematic energy band alignments at the interface and the corresponding current flow.
}\label{fig4}
\end{figure}
Despite faster carrier transit in Exf-$\mathrm{WS_2}$, CVD-$\mathrm{WS_2}$ produces a larger net photocurrent, qualitatively indicated by the THz signal amplitude in Fig. \ref{fig4}(a). 
The counterintuitive enhancement of the THz signal originates from asymmetric carrier-transfer at the rough CVD-$\mathrm{WS_2}$/Au interface, as confirmed by our transport model. 
Based on the net photocurrent direction indicated by the waveform of InSb in Fig. \ref{fig3}(d) and considerations of the interfacial band alignment, we model the carrier dynamics by simultaneous hot-electron and hot-hole transfer from $\mathrm{WS_2}$ to Au (see Section 6 of the Supplementary Information).
For CVD-$\mathrm{WS_2}$, the combination of $\tau_{\mathrm{tr,e}} = 135\ \mathrm{fs}$ and $\tau_{\mathrm{tr,h}} = 90\ \mathrm{fs}$ for electron and hole transport times, respectively, reproduces the detected THz waveform. 
Thus, the interfacial tunneling barrier in CVD-$\mathrm{WS_2}$ selectively favors hole transport, producing a pronounced hole-dominated net current (see Fig.\ref{fig4}(b)).
In contrast, the modeled response for uniform, strongly coupled contacts in exfoliated $\mathrm{WS_2}$/Au is obtained with $\tau_{\mathrm{tr,e}} = 60\ \mathrm{fs}$ and $\tau_{\mathrm{tr,h}} = 50\ \mathrm{fs}$. 
Under these conditions, electrons and holes transfer with comparable efficiencies, leading to near-complete cancellation of the photocurrent (Fig. \ref{fig4}(c)).
Notably, TR-ARPES studies of epitaxial TMDC monolayers on crystalline metal substrates have reported similarly symmetric and ultrafast carrier decay near the band edges  \cite{Cabo:2015, Ulstrup:2017}.
Observed saturation above $180\ \mathrm{\mu J/cm^2}$ in CVD-$\mathrm{WS_2}$ is attributed to space-charge effects (given a photo-excited carrier density of $\sim 4.3\cdot10^{13}\ \mathrm{cm^{-2}}$, photo-induced bleaching can be excluded). 
This saturation coincides with visible optical damage when measurements are conducted in ambient air, whereas Exf-$\mathrm{WS_2}$ shows no such damage, consistent with efficient charge transport.
The importance of charge separation for sample stability is discussed in the next section.
\section{Role of efficient charge carrier separation and extraction on photo-induced degradation}\label{sec4}
\begin{figure}[h]
\centering
\includegraphics[width=1\textwidth]{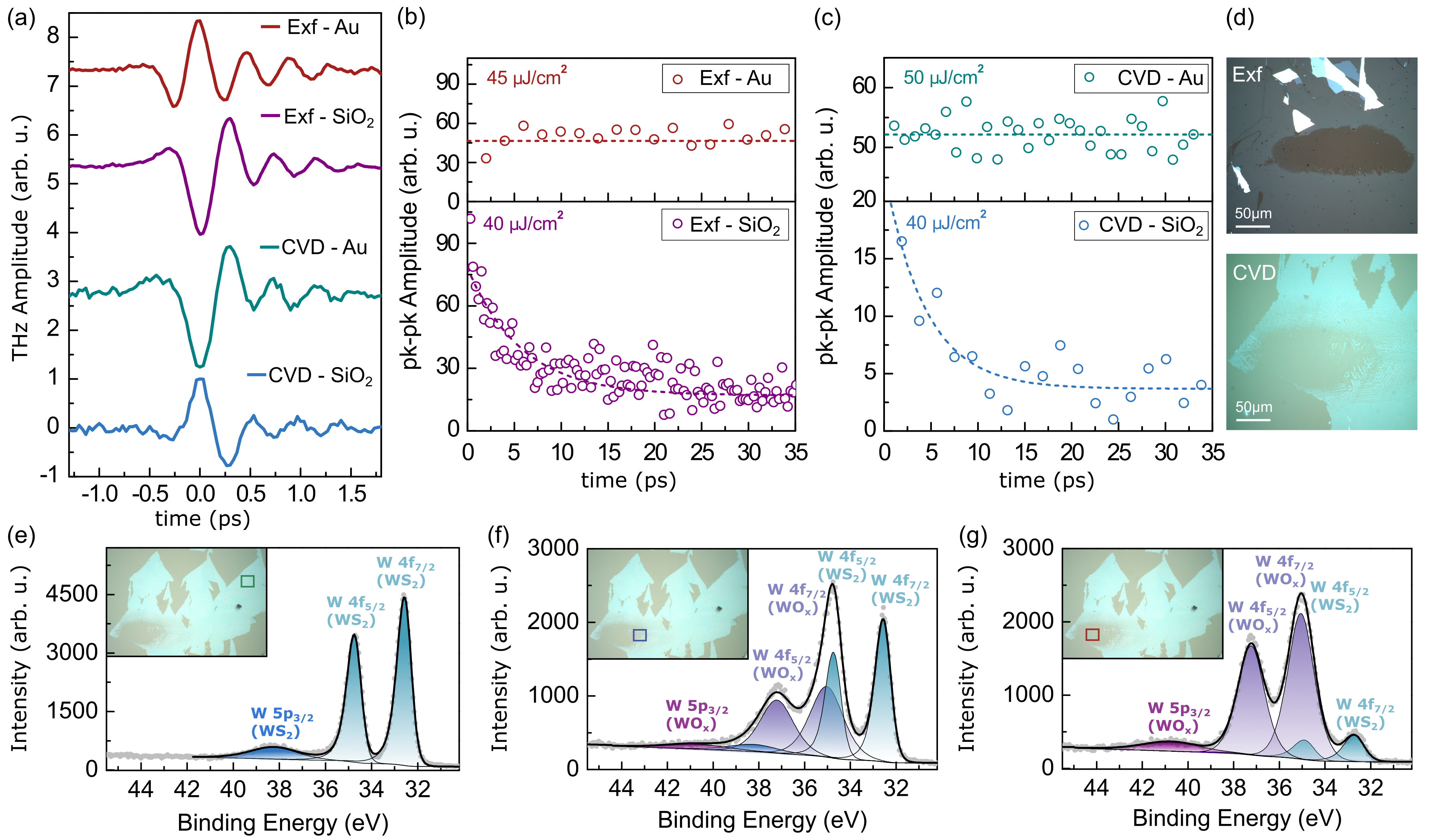}
\caption{(a) THz waveforms emitted from Exf-$\mathrm{WS_2}$ (red) and CVD-$\mathrm{WS_2}$ (green) on Au, as well as on $\mathrm{SiO_2}$ (purple and blue), respectively, measured with a 1 mm thick ZnTe crystal in dry air (0\% RH) following  $3.12\ \mathrm{eV}$ excitation. 
All samples exhibit stable THz emission when the chamber is purged with $\mathrm{N_2}$. 
(b)-(c) Peak-to peak THz amplitudes measured in ambient air as a function of time. 
$\mathrm{WS_2}$-Au maintain stable THz emission, whereas  signals from $\mathrm{WS_2}$-$\mathrm{SiO_2}$ samples decay within minutes. 
(d) Optical images showing clear photo-induced degradation of $\mathrm{WS_2}$-$\mathrm{SiO_2}$ samples after $3.17\ \mathrm{eV}$ excitation in ambient air containing $\mathrm{H_2O}$.
(e)-(g) XPS spectra of CVD-$\mathrm{WS_2}$ on $\mathrm{SiO_2}$ measured on (e) a pristine sample area without laser exposure, (f) an intermediately exposed area where optical damage starts to appear, (g) a significantly damaged area wherein fully oxidised $\mathrm{WS_2}$ appears transparent. The inset shows the position of the measurement.}\label{fig5}
\end{figure}

Many 2D materials are susceptible to oxidation and degradation under ambient conditions, making long-term stability and scalable protection strategies critical for practical applications. 
For ML $\mathrm{MoS_2}$ and $\mathrm{WS_2}$, oxidation typically initiates at grain boundaries or Sulfur vacancies after prolonged environmental exposure \cite{Kang:2017}, and is significantly accelerated by water acting as a catalytic agent \cite{Gao:2016, Ly:2014}.
More recently, photo-induced oxidation has been identified as the dominant degradation pathway for $\mathrm{WS_2}$ when illuminated with photon energies above the band gap \cite{Kotsakidis:2019}, yet the origin of the chemical reaction is not fully understood.  
To probe the role of charge transfer and separation for sample stability, we prepared ML $\mathrm{WS_2}$ on either Au or on insulating $\mathrm{SiO_2}$ substrates.
THz waveforms recorded after excitation with $3.12\ \mathrm{eV}$ pulses with a pulse fluence of $\sim50\ \mathrm{\mu J/cm^2}$ for both exfoliated and CVD-grown $\mathrm{WS_2}$ on Au and $\mathrm{SiO_2}$ are shown in Fig. \ref{fig5}(a). 
All samples exhibit stable THz emission in a $\mathrm{N_2}$ environment (0\% relative humidity), where oxidation chemistry is suppressed by the removal of $\mathrm{O_2}$ and $\mathrm{H_2O}$.
This observation excludes pure heating effects as the main mechanism for optical damage for the applied pump pulse energies.
In ambient air, THz emission from $\mathrm{WS_2}$/Au samples remains stable over an extended measurement time (see upper panels of Fig.\ref{fig5}(b)-(c)), while THz emission from $\mathrm{WS_2}$/$\mathrm{SiO_2}$ vanishes within minutes of excitation (see lower panels of Fig.\ref{fig5}(b)-(c)), accompanied by clear photo-induced damage visible under an optical microscope (Fig.\ref{fig5}(d)). 
X-ray photoelectron spectroscopy (XPS) confirms the near-complete, photo-induced oxidation of $\mathrm{WS_2}$ to optically transparent $\mathrm{WO_x}$ in the affected region. 
Figures \ref{fig5}(e)-(g) display the W 4f core-level spectra for (e) pristine, (f) adjacent, and (g) laser-damaged regions.
The pristine area (Fig. \ref{fig5}(e)) exhibits standard $\mathrm{WS_2}$ peaks ($\mathrm{W\ 4f_{7/2}}$ at $32.5\ \mathrm{eV}$ and $\mathrm{W\ 4f_{5/2}}$ at $34.7\ \mathrm{eV}$, bright blue curves) \cite{Morgan:2018}. 
Approaching the treated area (Figs. \ref{fig5}(f)-(g)), these components progressively diminish in intensity and shift to higher binding energies. 
This evolution, coupled with a growing $\mathrm{WO_x}$ contribution (purple curve), confirms the formation of tungsten oxide  \cite{Kieczka:2025}.\\
We conduct pump polarization dependent measurements on the emitted THz electric field to elucidate the THz generation mechanism for $\mathrm{WS_2}$ MLs on the dielectric substrate (see Fig. S8 of supplemental information).
While THz emission in $\mathrm{WS_2}$/Au samples is solely attributed to an out-of-plane interfacial charge transfer,
the polarization dependence of the THz field reveals an in-plane transient response in $\mathrm{WS_2}$/$\mathrm{SiO_2}$, associated with a resonant optical rectification process. 
This difference indicates that, in $\mathrm{WS_2}$/$\mathrm{SiO_2}$, photoexcited carriers remain confined within the ML and release excess energy through defect-assisted recombination, triggering oxidation and degradation. 
In contrast, efficient charge separation and extraction into Au on a sub-$100\ \mathrm{fs}$ timescale suppresses recombination-driven energy release, mitigating the onset of chemical reactions and stabilizing the ML.
Notably, when the excitation energy is in resonance with C-excitons or higher, only a few hot carriers relax toward the direct band edges resulting in a radiative recombination process \cite{Feng:2022}.
Due to the parallel band structure, the majority of C-excitons undergo self-separation, with electrons and holes relaxing to their immediate band extrema at the $\Lambda$ valley and $\Gamma$ crest via intraband cooling on a picosecond time scale  \cite{Kozawa:2014}.
Thus, ultrafast separation and extraction of photoexcited carriers \ \textemdash \ particularly in samples with strong interlayer coupling\ \textemdash \ provide a twofold stabilizing effect. 
They suppress the release of exciton binding energy, which can exceed 100 meV \cite{Wang_Gang:2018}, and they mitigate phonon-scattering pathways associated with the nested C-exciton band structure, thereby reducing heat accumulation in the monolayer.
Recently, Sim \textit{et al.} \cite{Sim:2026} reported on suppressed degradation in 8-layer $\mathrm{MoTe_2}$ on metal substrates, attributing it to interfacial charge transfer and energy band bending. They further propose that conductive substrates act as grounded carrier reservoirs enabling long-range dissipation of oxidation-induced charges. In contrast, in ML $\mathrm{WS_2}$ on Au \textemdash \ where energy band bending is not applicable \textemdash \ we identify ultrafast charge separation and the resulting suppression of recombination-driven energy release as the dominant protection mechanism. 
This is further supported by measurements on $\mathrm{MoS_2}$/$\mathrm{WS_2}$ heterostructures  as well as on samples with reversed stacking order ($\mathrm{WS_2}$/$\mathrm{MoS_2}$), both on $\mathrm{SiO_2}$ (see supplemental information Fig. S9).
Excitation above the band gap generates interlayer excitons \ \textemdash \ with holes transferring to $\mathrm{WS_2}$ and electrons to $\mathrm{MoS_2}$ \ \textemdash \  and a consequent transient photocurrent. 
Thus, depending on the stacking order, either electrons or holes remain near the surface, while charge separation reduces exciton recombination.
Under prolonged exposure in ambient air, the heterostructures  remain stable  \ \textemdash \ independent of the remaining charge carrier species at the sample surface \ \textemdash \ whereas adjacent $\mathrm{WS_2}$ monolayers degrade.\\
These findings demonstrate that interface engineering \ \textemdash \ specifically promoting ultrafast charge separation\ \textemdash \ can act as an intrinsic protection strategy, stabilizing 2D semiconductors without encapsulation and enabling scalable device architectures.
\section{Conclusion}\label{sec5}
By directly resolving ultrafast charge carrier dynamics in ML $\mathrm{WS_2}$/Au heterostructures, this work establishes a clear link between interface morphology, band alignment, directionality and efficiency of charge transfer, and  degradation pathways.
While samples with an intimate interface \ \textemdash \  causing orbital hybridization at the metal-TMDC junction \ \textemdash \  enable faster  transfer dynamics, heterostructures with a discontinuous contact produce a larger hole-dominated net photocurrent when excited above the band gap energy, due to asymmetric carrier flow across an interfacial tunnel barrier. 
Such results establish interfacial design as a central control parameter for tailoring both optoelectronic performance and long-term material robustness.
In particular, the identification of ultrafast charge separation and extraction as an intrinsic stabilization mechanism suggests a new strategy for mitigating photo-induced degradation without relying on encapsulation. 
This concept can be extended to engineered heterostructures and contact architectures that promote directional carrier flow, enabling devices that are simultaneously efficient and resilient under optical excitation.
\section{Methods}\label{sec6}
\textbf{Exf-$\mathbf{\mathrm{WS_2}}$ sample fabrication}: A 100 nm-thick Au film was deposited onto a silicon wafer, followed by the application of a UV-curable epoxy and bonding to a fused silica substrate. 
Upon curing, the Au film was mechanically lifted off, replicating the ultra-flat morphology of the underlying Si template.
AFM images of the Au substrate reveal an ultra-smooth surface with an average roughness of $0.251\ \mathrm{nm}$.
A freshly cleaved $\mathrm{WS_2}$ bulk crystal was then gently pressed onto the exposed Au surface. 
Subsequent lift-off of the bulk crystal left ML $\mathrm{WS_2}$ adhered to the Au film, facilitated by the strong adhesion between the Au and the ML. 
The procedure results in a flat Au-$\mathrm{WS_2}$ stack with a surface roughness of $0.255\ \mathrm{nm}$.\\
\textbf{CVD-$\mathrm{WS_2}$ sample fabrication}:
Monolayer $\mathrm{WS_2}$ was synthesized on sapphire substrates using the dip-coating hybrid metal-organic chemical vapor deposition (Hy-MOCVD) method, as described in our previous work \cite{Zhepeng:2024}. 
As-grown $\mathrm{WS_2}$ was transferred onto a gold substrate via a poly(methyl methacrylate) PMMA-assisted wet transfer process. 
Briefly, the sample was spin-coated with a PMMA layer and baked on a hot plate at $100^{\circ}\mathrm{C}$ for $5\ \mathrm{min}$. 
The PMMA/ $\mathbf{\mathrm{WS_2}}$ film was then delaminated in deionized (DI) water and transferred onto a $200\ \mathrm{nm}$ Au film deposited on a silica substrate via thermal evaporation. 
After transfer, the sample was dried by baking at $100^{\circ}\mathrm{C}$ for $15\ \mathrm{min}$, and the PMMA layer was subsequently removed by soaking in acetone for $15\ \mathrm{min}$.\\
\textbf{AFM and KPF Characterization:}
Atomic force microscopy (AFM) and Kelvin probe force microscopy (KPFM) measurements were performed using a Bruker Icon AFM system. 
AFM imaging was conducted in ScanAsyst mode using an NSC19 probe to evaluate the surface roughness of the samples. Surface potential 
measurements were carried out in Surface potential (AM-KPFM) mode using an NSC18Pt-coated probe. For KPFM measurements, the Drive 
2 routing was set to "Tip".\\
\textbf{Time-domain THz emission spectroscopy (TES)}:
For TES, a mode-locked Ti:sapphire laser emitting $795\ \mathrm{nm}$ pulses ($60\ \mathrm{fs}$, $5.12\ \mathrm{MHz}$ repetition) is employed.
A beam splitter divides the laser output into excitation and probe beams. 
For above band gap excitation, the fundamental pulse is frequency doubled in a $0.5\ \mathrm{mm}$ thick BBO crystal, yielding an estimated pulse duration of $76\ \mathrm{fs}$, accounting for group velocity dispersion.  
The excitation pulse is chopped at $320\ \mathrm{kHz}$ with an acousto optic modulator and focused onto the sample in reflection geometry. 
The emitted THz pulse in specular direction is collimated with a set of parabolic mirrors and focused onto an electro-optic sampling unit, comprising a nonlinear crystal (ZnTe or GaP), a quarter-wave plate, and a Wollaston prism. 
The THz wave induces birefringence in the crystal, altering the polarization of the probe pulse. 
The change in polarization creates an intensity asymmetry detected by a balanced photodetector and lock-in amplifier. 
By scanning the delay between the excitation and probe pulses, the THz waveform is recorded.\\
\textbf{X-ray photoelectron spectroscopy (XPS)}: 
XPS spectra were collected using a Physical Electronics PHI Versaprobe III with a hemispherical energy analyzer and a monochromatic aluminum K$\alpha$ X-ray source (1486.6 eV). 
Samples were attached to the sample stage through a copper clamp. 
Charging was prevented through the instrument's charge compensation system. 
Data were collected using a $50\ \mu m$, $12.5\ \mathrm{W}$ focused X-ray beam at a base pressure of $10^{-9}$ mbar, and a take-off angle of $45^{\circ}$. 
Survey scans were collected with a pass energy of $140.00\ \mathrm{eV}$ and a step size of $0.5\ \mathrm{eV}$. 
High-resolution scans of peaks of interest were collected with a pass energy of  $27.00\ \mathrm{eV}$ and a step size of $0.05\ \mathrm{eV}$. 
Data were analyzed with CasaXPS software. 
Peak fittings were performed using a Sh-Tougaard background. 
Gaussian Lorentzian product (GL) and Lorentzian Asymmetric (LA) lineshapes have been used for the peak fittings. 
W 4f components have been constrained with a $2.18\ \mathrm{eV}$ peak separation and to the same Full Width Half Maximum (FWHM). 
The areas were constrained with the 4:3 branching ratio. 
The $\mathrm{W\ 5p_{3/2}}$ component for each chemical phase was constrained to be $5.7\ \mathrm{eV}$ higher in binding energy than the respective phase of $\mathrm{W\ 4f_{7/2}}$ component. 
 $\mathrm{W\ 5p_{3/2}}$ has been constrained for a maximum FWHM of $2.5\ \mathrm{eV}$ for the sulfide component and $3.2\ \mathrm{eV}$ for the oxide component.
\backmatter

\bmhead{Supplementary information}

The article has accompanying supplementary information. 

\bmhead{Acknowledgements}
This work was supported by the Department of Energy, Office of Basic Energy Sciences, Division of Materials Sciences and Engineering, under contract DE-AC02-76SF00515, and funded in part by the Austrian Science Fund (FWF)[10.55776/J4721].
\section*{Declarations}
%
%
\bmhead{Competing interests}
On behalf of all authors, the corresponding author states that there is no conflict of interest.
\bmhead{Data availability}
All relevant data are available from the authors.

\noindent

\bibliography{NM_ref}

@article{Wang:2012,
author = {Wang, Qing Hua and Kalantar-Zadeh, Kourosh and Kis, Andras and Coleman, Jonathan N. and Strano, Michael S.},
title = {{Electronics and optoelectronics of two-dimensional transition metal dichalcogenides}},
journal = {Nature Nanotechnology},
volume = {7},
number = {11},
pages = {699--712},
year = {2012},
doi = {10.1038/nnano.2012.193},
URL = {https://doi.org/10.1038/nnano.2012.193},
}

@Article{Schmidt:2015,
author = {Schmidt, Hennrik and Giustiniano, Francesco and Eda, Goki},
title  = {{Electronic transport properties of transition metal dichalcogenide field-effect devices: surface and interface effects}},
journal  = {Chem. Soc. Rev.},
year  = {2015},
volume  = {44},
issue  = {21},
pages  = {7715-7736},
publisher  = {The Royal Society of Chemistry},
doi  = {10.1039/C5CS00275C},
url  = {http://dx.doi.org/10.1039/C5CS00275C},
}

@Article{Liao:2020,
author = {Liao, Wugang and Zhao, Siwen and Li, Feng and Wang, Cong and Ge, Yanqi and Wang, Huide and Wang, Shibo and Zhang, Han},
title  = {{Interface engineering of two-dimensional transition metal dichalcogenides towards next-generation electronic devices: recent advances and challenges}},
journal  = {Nanoscale Horiz.},
year  = {2020},
volume  = {5},
issue  = {5},
pages  = {787-807},
publisher  = {The Royal Society of Chemistry},
doi  = {10.1039/C9NH00743A},
url  = {http://dx.doi.org/10.1039/C9NH00743A},
}

@Article{Schaibley:2016,
author = {Schaibley, John R. and Yu, Hongyi and Clark, Genevieve and Rivera, Pasqual and Ross, Jason S. and Seyler, Kyle L. and Yao, Wang and Xu, Xiaodong},
title  = {{Valleytronics in 2D materials}},
journal  = {Nature Reviews Materials},
year  = {2016},
volume  = {1},
issue  = {11},
pages  = {16055},
doi  = {10.1038/natrevmats.2016.55},
url  = {https://doi.org/10.1038/natrevmats.2016.55},
}

@article{Fang:2020,
author = {Fang Liu  and Wenjing Wu  and Yusong Bai  and Sang Hoon Chae  and Qiuyang Li  and Jue Wang  and James Hone  and X.-Y. Zhu },
title = {{Disassembling 2D van der Waals crystals into macroscopic monolayers and reassembling into artificial lattices}},
journal = {Science},
volume = {367},
number = {6480},
pages = {903-906},
year = {2020},
doi = {10.1126/science.aba1416},
URL = {https://www.science.org/doi/abs/10.1126/science.aba1416},
}

@article{Zhepeng:2024,
author = {Zhang, Zhepeng and Hoang, Lauren and Hocking, Marisa and Peng, Zhenghan and Hu, Jenny and Zaborski, Gregory Jr. and Reddy, Pooja D. and Dollard, Johnny and Goldhaber-Gordon, David and Heinz, Tony F. and Pop, Eric and Mannix, Andrew J.},
title = {{Chemically Tailored Growth of 2D Semiconductors via Hybrid Metal–Organic Chemical Vapor Deposition}},
journal = {ACS Nano},
volume = {18},
number = {37},
pages = {25414-25424},
year = {2024},
doi = {10.1021/acsnano.4c02164},
URL = {https://doi.org/10.1021/acsnano.4c02164},
}

@article{Pollmann:2021,
author = {Pollmann, Erik and Sleziona, Stephan and Foller, Tobias and Hagemann, Ulrich and Gorynski, Claudia and Petri, Oliver and Madauß, Lukas and Breuer, Lars and Schleberger, Marika},
title = {{Large-Area, Two-Dimensional MoS2 Exfoliated on Gold: Direct Experimental Access to the Metal–Semiconductor Interface}},
journal = {ACS Omega},
volume = {6},
number = {24},
pages = {15929-15939},
year = {2021},
doi = {10.1021/acsomega.1c01570},
URL = {https://doi.org/10.1021/acsomega.1c01570},
}

@article{Gong:2014,
author = {Gong, Cheng and Colombo, Luigi and Wallace, Robert M. and Cho, Kyeongjae},
title = {{The Unusual Mechanism of Partial Fermi Level Pinning at Metal–MoS2 Interfaces}},
journal = {Nano Letters},
volume = {14},
number = {4},
pages = {1714-1720},
year = {2014},
doi = {10.1021/nl403465v},
URL = {https://doi.org/10.1021/nl403465v},
}

@article{Markeev:2021,
author = {Markeev, Pavel A. and Najafidehaghani, Emad and Gan, Ziyang and Sotthewes, Kai and George, Antony and Turchanin, Andrey and de Jong, Michel P.},
title = {{Energy-Level Alignment at Interfaces between Transition-Metal Dichalcogenide Monolayers and Metal Electrodes Studied with Kelvin Probe Force Microscopy}},
journal = {The Journal of Physical Chemistry C},
volume = {125},
number = {24},
pages = {13551-13559},
year = {2021},
doi = {10.1021/acs.jpcc.1c01612},
URL = {https://doi.org/10.1021/acs.jpcc.1c01612},
}

@article{Krolikowski:1970,
  title = {{Photoemission Studies of the Noble Metals. II. Gold}},
  author = {Krolikowski, W. F. and Spicer, W. E.},
  journal = {Phys. Rev. B},
  volume = {1},
  issue = {2},
  pages = {478--487},
  numpages = {0},
  year = {1970},
  month = {Jan},
  publisher = {American Physical Society},
  doi = {10.1103/PhysRevB.1.478},
  url = {https://link.aps.org/doi/10.1103/PhysRevB.1.478}
}

@article{Tang:2019,
author = {Tang, Hao and Shi, Bowen and Pan, Yuanyuan and Li, Jingzhen and Zhang, Xiuying and Yan, Jiahuan and Liu, Shiqi and Yang, Jie and Xu, Lianqiang and Yang, Jinbo and Wu, Mingbo and Lu, Jing},
title = {{Schottky Contact in Monolayer WS2 Field-Effect Transistors}},
journal = {Advanced Theory and Simulations},
volume = {2},
number = {5},
pages = {1900001},
doi = {https://doi.org/10.1002/adts.201900001},
url = {https://advanced.onlinelibrary.wiley.com/doi/abs/10.1002/adts.201900001},
year = {2019}
}

@article{Hill:2016,
author = {Hill, Heather M. and Rigosi, Albert F. and Rim, Kwang Taeg and Flynn, George W. and Heinz, Tony F.},
title = {{Band Alignment in MoS2/WS2 Transition Metal Dichalcogenide Heterostructures Probed by Scanning Tunneling Microscopy and Spectroscopy}},
journal = {Nano Letters},
volume = {16},
number = {8},
pages = {4831-4837},
year = {2016},
doi = {10.1021/acs.nanolett.6b01007}, 
URL = {https://doi.org/10.1021/acs.nanolett.6b01007},
}

@article{Dreesen:2002,
author = {Dreesen, L. and Humbert, C. and Celebi, M. and Lemaire, J.J. and Mani, A.A. and Thiry, P.A. and Peremans, A.},
title = {{Influence of the metal electronic properties on the sum-frequency generation spectra of dodecanethiol self-assembled monolayers on Pt(111), Ag(111) and Au(111) single crystals}},
journal = {Applied Physics B},
volume = {74},
number = {7},
pages = {621--625},
year = {2002},
doi = {10.1007/s00340-002-0924-6}, 
URL = {https://doi.org/10.1007/s00340-002-0924-6},
}

@article{Mohl:2020,
author = {Mohl, Melinda and Rautio, Anne-Riikka and Asres, Georgies Alene and Wasala, Milinda and Patil, Prasanna Dnyaneshwar and Talapatra, Saikat and Kordas, Krisztian},
title = {{2D Tungsten Chalcogenides: Synthesis, Properties and Applications}},
journal = {Advanced Materials Interfaces},
volume = {7},
number = {13},
pages = {2000002},
doi = {https://doi.org/10.1002/admi.202000002},
url = {https://advanced.onlinelibrary.wiley.com/doi/abs/10.1002/admi.202000002},
year = {2020},
}

@article{Kang:2013,
    author = {Kang, Jun and Tongay, Sefaattin and Zhou, Jian and Li, Jingbo and Wu, Junqiao},
    title = {{Band offsets and heterostructures of two-dimensional semiconductors}},
    journal = {Applied Physics Letters},
    volume = {102},
    number = {1},
    pages = {012111},
    year = {2013},
    month = {01},
    issn = {0003-6951},
    doi = {10.1063/1.4774090},
    url = {https://doi.org/10.1063/1.4774090},
}

@article{Frisenda:2018,
author = {Frisenda, Riccardo and Molina-Mendoza, Aday J. and Mueller, Thomas and Castellanos-Gomez, Andres and van der Zant, Herre S. J.},
title  = {{Atomically thin p–n junctions based on two-dimensional materials}},
journal  = {Chem. Soc. Rev.},
year  = {2018},
volume  = {47},
issue  = {9},
pages  = {3339-3358},
publisher  = {The Royal Society of Chemistry},
doi  = {10.1039/C7CS00880E},
url  = {http://dx.doi.org/10.1039/C7CS00880E},
}

@article{Liu:2016,
author = {Yuanyue Liu  and Paul Stradins  and Su-Huai Wei },
title = {{Van der Waals metal-semiconductor junction: Weak Fermi level pinning enables effective tuning of Schottky barrier}},
journal = {Science Advances},
volume = {2},
number = {4},
pages = {e1600069},
year = {2016},
doi = {10.1126/sciadv.1600069},
URL = {https://www.science.org/doi/abs/10.1126/sciadv.1600069},
}

@article{Eroglu:2021,
AUTHOR = {Ezgi Eroglu, Zeynep and Contreras, Dillon and Bahrami, Pouya and Azam, Nurul and Mahjouri-Samani, Masoud and Boulesbaa, Abdelaziz},
TITLE = {{Filling Exciton Trap-States in Two-Dimensional Tungsten Disulfide (WS2) and Diselenide (WSe2) Monolayers}},
JOURNAL = {Nanomaterials},
VOLUME = {11},
YEAR = {2021},
NUMBER = {3},
ARTICLE-NUMBER = {770},
URL = {https://www.mdpi.com/2079-4991/11/3/770},
PubMedID = {33803656},
ISSN = {2079-4991},
DOI = {10.3390/nano11030770}
}

@article{Ansari:2018,
author = {Narges Ansari and Farinaz Ghorbani},
journal = {J. Opt. Soc. Am. B},
number = {5},
pages = {1179--1185},
publisher = {Optica Publishing Group},
title = {{Light absorption optimization in two-dimensional transition metal dichalcogenide van der Waals heterostructures}},
volume = {35},
month = {May},
year = {2018},
url = {https://opg.optica.org/josab/abstract.cfm?URI=josab-35-5-1179},
doi = {10.1364/JOSAB.35.001179},
}

@article{Li:2014,
  title = {{Measurement of the optical dielectric function of monolayer transition-metal dichalcogenides: ${\mathrm{MoS}}_{2}$, $\mathrm{Mo}\mathrm{S}{\mathrm{e}}_{2}$, ${\mathrm{WS}}_{2}$, and $\mathrm{WS}{\mathrm{e}}_{2}$}},
  author = {Li, Yilei and Chernikov, Alexey and Zhang, Xian and Rigosi, Albert and Hill, Heather M. and van der Zande, Arend M. and Chenet, Daniel A. and Shih, En-Min and Hone, James and Heinz, Tony F.},
  journal = {Phys. Rev. B},
  volume = {90},
  issue = {20},
  pages = {205422},
  numpages = {6},
  year = {2014},
  month = {Nov},
  publisher = {American Physical Society},
  doi = {10.1103/PhysRevB.90.205422},
  url = {https://link.aps.org/doi/10.1103/PhysRevB.90.205422}
}

@article{Ulstrup:2017,
  title = {{Spin and valley control of free carriers in single-layer ${\mathrm{WS}}_{2}$}},
  author = {Ulstrup, S\o{}ren and \ifmmode \check{C}\else \v{C}\fi{}abo, Antonija Grubi\ifmmode \check{s}\else \v{s}\fi{}i\ifmmode \acute{c}\else \'{c}\fi{} and Biswas, Deepnarayan and Riley, Jonathon M. and Dendzik, Maciej and Sanders, Charlotte E. and Bianchi, Marco and Cacho, Cephise and Matselyukh, Dan and Chapman, Richard T. and Springate, Emma and King, Phil D. C. and Miwa, Jill A. and Hofmann, Philip},
  journal = {Phys. Rev. B},
  volume = {95},
  issue = {4},
  pages = {041405},
  numpages = {5},
  year = {2017},
  month = {Jan},
  publisher = {American Physical Society},
  doi = {10.1103/PhysRevB.95.041405},
  url = {https://link.aps.org/doi/10.1103/PhysRevB.95.041405}
}

@article{Cabo:2015,
author = {Grubišić Čabo, Antonija and Miwa, Jill A. and Grønborg, Signe S. and Riley, Jonathon M. and Johannsen, Jens C. and Cacho, Cephise and Alexander, Oliver and Chapman, Richard T. and Springate, Emma and Grioni, Marco and Lauritsen, Jeppe V. and King, Phil D. C. and Hofmann, Philip and Ulstrup, Søren},
title = {{Observation of Ultrafast Free Carrier Dynamics in Single Layer MoS2}},
journal = {Nano Letters},
volume = {15},
number = {9},
pages = {5883-5887},
year = {2015},
doi = {10.1021/acs.nanolett.5b01967},
note ={PMID: 26315566},
URL = {https://doi.org/10.1021/acs.nanolett.5b01967},
}

@article{Liu:2025,
author = {Liu, Xiongfang and Xing, Kaijian and Tang, Chi Sin and Sun, Shuo and Chen, Pan and Qi, Dong-Chen and Breese, Matk B.H. and Fuhrer, Michael S. and Wee, Andrew T.S.},
title = {{Contact resistance and interfacial engineering: Advances in high-performance 2D-TMD based devices}},
journal = {Progress in Materials Science},
volume = {148},
pages = {101390},
year = {2025},
doi = {doi.org/10.1016/j.pmatsci.2024.101390},
URL = {https://doi.org/10.1016/j.pmatsci.2024.101390},
}

@article{Shan:2019,
author = {Shan, Hangyong and Yu, Ying and Wang, Xingli and Luo, Yang and Du, Bowen and Han, Tianyang and Li, Bowen and Li, Yu and Wu, Jiarui and Lin, Feng and Shi, Kebin and Tay, Beng Kang and Liu, Zheng and Zhu, Xing and Fang, Zheyu},
title = {{Direct observation of ultrafast plasmonic hot electron transfer in the strong coupling regime}},
journal = {Light: Science \& Applications},
volume = {8},
 issue = {1},
pages = {9},
year = {2019},
doi = {10.1038/s41377-019-0121-6},
URL = {https://doi.org/10.1038/s41377-019-0121-6},
}

@article{Hong:2025,
author = {Chengyun Hong  and Hyundong Kim  and Ye Tao  and Jong Hyeon Lim  and Jin Yong Lee  and Ji-Hee Kim },
title = {{Ultrafast hot carrier extraction and diffusion at the $\mathrm{MoS_2}$/Au van der Waals electrode interface}},
journal = {Science Advances},
volume = {11},
number = {1},
pages = {eadr1534},
year = {2025},
doi = {10.1126/sciadv.adr1534},
URL = {https://www.science.org/doi/abs/10.1126/sciadv.adr1534},
}

@article{Kang:2017,
author = {Kang, Kyungnam and Godin, Kyle and Kim, Young Duck and Fu, Shichen and Cha, Wujoon and Hone, James and Yang, Eui-Hyeok},
title = {{Graphene-Assisted Antioxidation of Tungsten Disulfide Monolayers: Substrate and Electric-Field Effect}},
journal = {Advanced Materials},
volume = {29},
number = {18},
pages = {1603898},
doi = {https://doi.org/10.1002/adma.201603898},
url = {https://advanced.onlinelibrary.wiley.com/doi/abs/10.1002/adma.201603898},
year = {2017},
}

@article{Gao:2016,
author = {Gao, Jian and Li, Baichang and Tan, Jiawei and Chow, Phil and Lu, Toh-Ming and Koratkar, Nikhil},
title = {{Aging of Transition Metal Dichalcogenide Monolayers}},
journal = {ACS Nano},
volume = {10},
number = {2},
pages = {2628-2635},
year = {2016},
doi = {10.1021/acsnano.5b07677},
URL = { https://doi.org/10.1021/acsnano.5b07677},
}

@article{Ly:2014,
author = {Ly, Thuc Hue and Chiu, Ming-Hui and Li, Ming-Yang and Zhao, Jiong and Perello, David J. and Cichocka, Magdalena Ola and Oh, Hye Min and Chae, Sang Hoon and Jeong, Hye Yun and Yao, Fei and Li, Lain-Jong and Lee, Young Hee},
title = {{Observing Grain Boundaries in CVD-Grown Monolayer Transition Metal Dichalcogenides}},
journal = {ACS Nano},
volume = {8},
number = {11},
pages = {11401-11408},
year = {2014},
doi = {10.1021/nn504470q},
URL = {https://doi.org/10.1021/nn504470q},
}

@article{Kotsakidis:2019,
author = {Kotsakidis, Jimmy C. and Zhang, Quianhui and Vazquez de Parga, Amadeo L. and Currie, Marc and Helmerson, Kristian and Gaskill, D. Kurt and Fuhrer, Michael S.},
title = {{Oxidation of Monolayer WS2 in Ambient Is a Photoinduced Process}},
journal = {Nano Letters},
volume = {19},
number = {8},
pages = {5205-5215},
year = {2019},
doi = {10.1021/acs.nanolett.9b01599},
URL = {https://doi.org/10.1021/acs.nanolett.9b01599},
}

@article{Feng:2022,
author = {Feng, Jiying and Li, Yuanzheng and Li, Jixiu and Feng, Qiushi and Xin, Wei and Liu, Weizhen and Xu, Haiyang and Liu, Yichun},
title = {{Engineering Relaxation-Paths of C-Exciton for Constructing Band Nesting Bypass in WS2 Monolayer}},
journal = {Nano Letters},
volume = {22},
number = {9},
pages = {3699-3706},
year = {2022},
doi = {10.1021/acs.nanolett.2c00509},
note ={PMID: 35481760},
URL = {https://doi.org/10.1021/acs.nanolett.2c00509},
eprint = { https://doi.org/10.1021/acs.nanolett.2c00509}
}

@article{Kozawa:2014,
author = {Kozawa, Daichi and Kumar, Rajeev and Carvalho, Alexandra and Kumar Amara, Kiran and Zhao, Weijie and Wang, Shunfeng and Toh, Minglin and Ribeiro, Ricardo M. and Castro Neto, A. H. and Matsuda, Kazunari and Eda, Goki},
title = {{Photocarrier relaxation pathway in two-dimensional semiconducting transition metal dichalcogenides}},
journal = {Nature Communications},
volume = {5},
number = {1},
pages = {4543},
doi = {10.1038/ncomms5543},
url = {https://doi.org/10.1038/ncomms5543},
year = {2014},
}

@article{Wang_Gang:2018,
  title = {{Colloquium: Excitons in atomically thin transition metal dichalcogenides}},
  author = {Wang, Gang and Chernikov, Alexey and Glazov, Mikhail M. and Heinz, Tony F. and Marie, Xavier and Amand, Thierry and Urbaszek, Bernhard},
  journal = {Rev. Mod. Phys.},
  volume = {90},
  issue = {2},
  pages = {021001},
  numpages = {25},
  year = {2018},
  month = {Apr},
  publisher = {American Physical Society},
  doi = {10.1103/RevModPhys.90.021001},
  url = {https://link.aps.org/doi/10.1103/RevModPhys.90.021001}
}

@article{Chen:2016,
url = {https://doi.org/10.1038/ncomms12512},
title = {{Ultrafast formation of interlayer hot excitons in atomically thin MoS2/WS2 heterostructures}},
author = {Chen, Hailong and Wen, Xiewen and Zhang, Jing and Wu, Tianmin and Gong, Yongji and Zhang, Xiang and Yuan, Jiangtan and Yi, Chongyue and Lou, Jun and Ajayan, Pulickel M. and Zhuang, Wei and Zhang, Guangyu and Zheng, Junrong},
pages = {12512},
volume = {7},
number = {1},
journal = {Nature Communications},
doi = {10.1038/ncomms12512},
year = {2016},
}

@article{Ma:2019,
author = {Eric Yue Ma  and Burak Guzelturk  and Guoqing Li  and Linyou Cao  and Zhi-Xun Shen  and Aaron M. Lindenberg  and Tony F. Heinz },
title = {{Recording interfacial currents on the subnanometer length and femtosecond time scale by terahertz emission}},
journal = {Science Advances},
volume = {5},
number = {2},
pages = {eaau0073},
year = {2019},
doi = {10.1126/sciadv.aau0073},
URL = {https://www.science.org/doi/abs/10.1126/sciadv.aau0073},
eprint = {https://www.science.org/doi/pdf/10.1126/sciadv.aau0073},
}

@article{Morgan:2018,
    author = {Morgan, David J.},
    title = {{Core-level spectra of powdered tungsten disulfide, WS2}},
    journal = {Surface Science Spectra},
    volume = {25},
    number = {1},
    pages = {014002},
    year = {2018},
    month = {07},
    issn = {1055-5269},
    doi = {10.1116/1.5030093},
    url = {https://doi.org/10.1116/1.5030093},
}

@Article{Kieczka:2025,
author ="Kieczka, Daria and Bussolotti, Fabio and Maddumapatabandi, Thathsara D. and Bosman, Michel and Shluger, Alexander and Regoutz, Anna and Goh, Kuan Eng Johnson",
title  ={{Unveiling surface dynamics: in situ oxidation of defective WS2}},
journal  ="Nanoscale",
year  ="2025",
volume  ="17",
issue  ="16",
pages  ="10082-10094",
publisher  ="The Royal Society of Chemistry",
doi  ="10.1039/D4NR04992F",
url  ="http://dx.doi.org/10.1039/D4NR04992F",
}

@article{Xiaochi:2022,
author = {Liu, Xiaochi and Choi, Min Sup and Hwang, Euyheon and Yoo, Won Jong and Sun, Jian},
title = {Fermi Level Pinning Dependent 2D Semiconductor Devices: Challenges and Prospects},
journal = {Advanced Materials},
volume = {34},
number = {15},
pages = {2108425},
doi = {https://doi.org/10.1002/adma.202108425},
url = {https://advanced.onlinelibrary.wiley.com/doi/abs/10.1002/adma.202108425},
eprint = {https://advanced.onlinelibrary.wiley.com/doi/pdf/10.1002/adma.202108425},
year = {2022}
}

@article{Wang_Yan:2019,
author = {Wang, Yan and Kim, Jong Chan and Wu, Ryan J. and Martinez, Jenny and Song, Xiuju and Yang, Jieun and Zhao, Fang and Mkhoyan, Andre and Jeong, Hu Young and Chhowalla, Manish},
title = {Van der Waals contacts between three-dimensional metals and two-dimensional semiconductors},
journal = {Nature},
volume = {568},
issue = {7750},
pages = {70-74},
doi = {10.1038/s41586-019-1052-3},
url = {https://doi.org/10.1038/s41586-019-1052-3},
year = {2019}
}

@article{Sim:2026,
author = {Sim, Yeoseon and Kim, Se-Yang and Park, Soon-Dong and Lee, Hyeonwoo and Kim, Junghwa and Jang, Sora and Wang, Jaewon and Song, Seunguk and Kwak, Jinsung and Lee, Zonghoon and Jeong, Changwook and Kim, Sung Youb and Kwon, Soon-Yong},
title = {Interfacial Control of Degradation Pathways in 2D Heterostructures},
journal = {Advanced Functional Materials},
volume = {36},
number = {2},
pages = {e16434},
doi = {https://doi.org/10.1002/adfm.202516434},
url = {https://advanced.onlinelibrary.wiley.com/doi/abs/10.1002/adfm.202516434},
year = {2026}
}

@article{Gollner:2026,
author = {Gollner, Claudia and Lindenberg, Aaron and Heinz, Tony F.},
title = {{Time-domain terahertz emission spectroscopy on van der Waals materials}},
journal = {MRS Communications},
doi = {10.1557/s43579-026-00940-z},
url = {https://doi.org/10.1557/s43579-026-00940-z},
year = {2026}
}

\end{document}